\documentclass[traditabstract]{aa}

\usepackage{graphicx}
\usepackage{natbib}

\title{A Wave Scattering Theory of Solar Seismic Power Haloes}
\author{Shravan M. Hanasoge}
\institute{Max-Planck-Institut-f\"{u}r-Sonnensystemforschung, 37191 Katlenburg-Lindau, Germany}
\titlerunning{Haloes by Wave Scattering}
\authorrunning{Hanasoge}
\date{Received 7 May 2009\\
Accepted 23 June 2009}
\abstract{
Spatial maps of the high-pass frequency filtered time-averaged root-mean-squared (RMS) Doppler velocities
tend to show substantial decrements within regions of strong field and curiously, randomly distributed patches of enhancement in the vicinity.
We propose that these haloes or enhancements are a consequence of magnetic-field-induced mode mixing (scattering), resulting in the 
preferential powering of waves that possess strong surface velocity signatures (i.e. scattering from low to high wavenumbers). Evidently, this process can occur in the reverse,  and therefore in order
to determine if the haloes are indeed caused by mode mixing, we must answer the question: {\it how are acoustic waves scattered by magnetic fields?} Through 
simulations of the interactions between waves and sunspots and models of plage, we demonstrate that the high to low modal order scattering channels are favoured. 
With increasing frequency and consequently, decreasing wavelength, a growing number of modes are scattered by the sunspot, thereby rendering the enhancements most visible 
around the high-frequency parts of the spectrum. The haloes obtained from the simulations are on the same order of magnitude but weaker than those observed. We also present observational 
evidence to support this theory: observations of active region AR9787 are firstly frequency filtered to isolate the 5-6 mHz signals and secondly, decomposed into three wavenumber bandpasses, 
$l - [0,400], [400,800], [800,2222]$. With increasing wavenumber, the extent of the halo effect is seen increase dramatically, in line with theoretical expectation.}

\begin{document}
\maketitle
\thanks{E-mail:hanasoge@mps.mpg.de}


\section{Introduction}
Observations of acoustic power in regions of strong magnetic field are numerous: \citet{Lites,Brown92,Braun92,bogdan93,balthasar,hindman98,donea00,Jain2002,moretti,nagashima}.
In general, there is agreement that trapped acoustic wave power (with frequencies below the acoustic cut-off) measured using Doppler velocities is strongly suppressed in 
the strongest field regions, and that in the weak field surrounding area, there is an enhancement of wave power at frequencies comparable to and above the acoustic cut-off. 
Corresponding measurements in intensity however do not quite provide such a consistent picture; numerous differences are seen in the structure and spatial locations of the haloes.
The variety of instruments and optical spectral lines used in taking these observations may be the source of the discrepancies \citep[e.g.][]{moretti}. Furthermore, because intensity is a complex metric,
and interpreting changes thereof can be non-trivial, we stick to the Doppler measurements. 

There have been several theories attempting to explain the emergence of the haloes. \citet{Brown92, donea00} subscribe to the hypothesis of excess wave source activity in the vicinity
of active regions. Such an assumption is entirely plausible given the remarkable associated changes in the turbulent layers of the solar photosphere. More recently, \citet{jacoutot2008},
based on numerical simulations of magnetoconvection found that with increasing strength of network magnetic fields, the oscillation spectrum shifted to higher frequencies. 
Alternately, \citet{kuridze2008} and \citet{hanasoge_mag} have proposed that a linear (i.e. small wave amplitudes) magnetohydrodynamic (MHD) wave interaction mechanism 
may be responsible for the haloes. While \citet{hanasoge_mag}, based on numerical simulations of MHD wave propagation, merely speculated about the existence of such a mechanism, \citet{kuridze2008}
suggested that overlying canopy magnetic fields may locally raise the acoustic cut-off frequency, thereby trapping the high frequency waves. On a related note, the observations of 
\citet{Braun2000} indicate that high frequency waves undergo enhanced reflections in active regions.

Continuing on the path set forth in \citet{hanasoge_mag}, we posit here that wave scattering induces a transfer of energies across modes at constant frequency, a consequence of
which is the appearance of the velocity amplitude enhancements in the vicinity of active regions. Very recently, \citet{Gordovskyy2009}, in a study of $p$-mode
 oscillation amplitude anomalies around active regions \citep[e.g.][]{hindman97} reached a similar conclusion, i.e., that the scattering occurs preferentially from high to low radial orders.
In this contribution, we dynamically solve the linearized three-dimensional MHD equations in order to examine the mode interactions with strong-field flux tubes and plage. The principal assumption in these
calculations is that of linearity, or low wave amplitudes in comparison to the local phase and group velocities.  Analyses of the simulated data display signs of biased scattering and the consequent 
emergence of velocity enhancements. We also present observational evidence to support this theory.



\section{Theory}
 A wave propagating in the quiet Sun has an associated displacement (for example) eigenfunction. When this wave encounters an anomaly, the wave displacements may no longer be well represented 
 by just the quiet Sun eigenfunction, but rather, require an additional basis of eigenfunctions. These additional eigenfunctions can only be introduced into the system by the creation of a corresponding set of 
 waves. The energy required to produce these additional or scattered waves is extracted from the incident wave, resulting in an energy loss of the original wave. Due to the low amplitudes of the seismic 
 waves of relevance to helioseismology, we may treat scattering in the Sun as a linear phenomenon. Furthermore, the background medium and associated scatterers are considered to be time stationary. 
 Consequently, wave interactions with scatterers can only result in the redistribution of energies across wavenumbers at constant frequency.

\begin{figure}
\includegraphics[width=\linewidth]{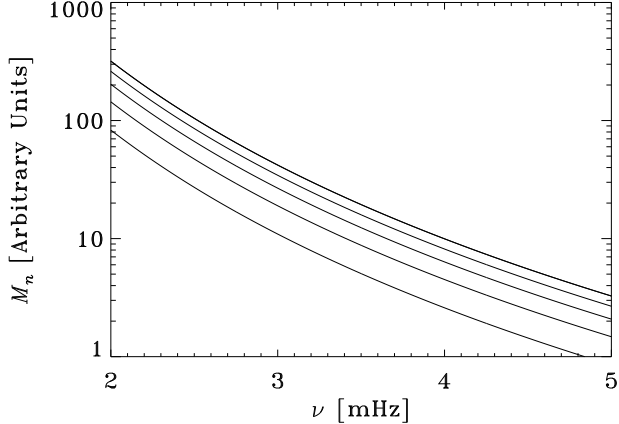}
\caption{Arbitrarily scaled mode mass ($M_n$) as a function of frequency for different $p_n$ ridges computed for a polytrope \citep[equation (3.13) of][]{bogdan96}. The curves from bottom to top 
correspond to the $n=1,..,5$ $p_n$ ridges. The
ratios of the masses are useful indicators of the degree of change in the surface velocity signatures that can be effected by the scatter from one mode to another.
\label{modemass}}
\end{figure}

The velocities observed at the solar surface are a sum of the velocities of a large number of modes, each of which has a characteristic surface amplitude. The ratio of the time-average energy of a mode 
to the surface mean square velocity is termed {\it mode mass} \citep[e.g.][]{bogdan96}; low mode mass waves possess strong surface signatures and conversely. Mode mass 
is a monotonically decreasing function of wavenumber and frequency, as seen in Figure~\ref{modemass}. Consider, for example, a $p_4$ mode of unit energy being fully transformed 
into $p_1$ at a specific frequency. From Figure 1, we see that the $p_4$ to $p_1$ mode mass ratio 
is approximately 5, indicating the magnitude by which the surface velocity square would increase were such a transformation to occur. A small amount of scatter can thus result in significant changes in the observed surface Doppler
velocities. 

Observations of the quiet Sun photosphere show a modal velocity spectrum that peaks around a frequency of $\nu \sim 3.6$ mHz. This is the consequence of an impasse reached between damping, emission, and 
mode mass. The presence of strong scatterers can disturb this delicate equilibrium and reconfigure the spectrum. This,
we propose, is precisely what is happening in the vicinity of active regions. High mode mass waves are depleted of their energies, which are then channeled into the low mode mass counterparts. 

\begin{figure}
\includegraphics[width=\linewidth]{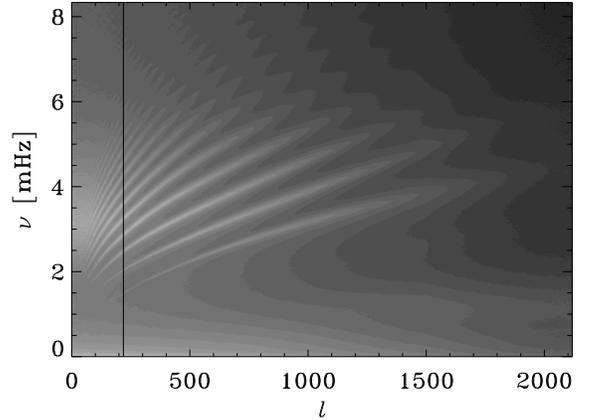}
\caption{Log power spectrum from 1440 minutes of full disk observations of the active region AR9787 by the Michelson Doppler Imager instrument \citep[MDI;][]{scherrer95}. The 
frequency is denoted by $\nu$ and the spherical harmonic degree by $l = 2\pi R_\odot/\lambda$, where $\lambda$ is the wavelength. 
 The vertical line marks $2\pi R_\odot/l_c$, where $l_c = 20$ Mm, the characteristic 
length scale of a hypothetical scatterer. We may broadly classify the scattering process into two regimes, weak and strong, on the left and right sides of the line respectively. Roughly speaking, modes on the left of the line
do not ``see" the scatterer very well because of their relatively-large wavelengths. At high frequencies, the wavelength decreases and many more modes couple with the scatterer, potentially significantly altering 
the distribution of energies with wavenumber. Note that at 3 mHz, four ridges lie in the ``strong'' scattering regime whereas at 5 mHz, 11 radial orders fall in this regime.
\label{interact}}
\end{figure}

Consider a hypothetical situation where waves are
scattered by a flux tube of size 15 Mm or so. Waves with wavelengths much larger than this flux tube do not couple with it very strongly and may be classified as being weakly scattered. Conversely, smaller
wavelengths (also higher frequencies) are strongly scattered. It is therefore reasonable to conclude that with increasing frequency, more modes fall into the strong scattering regime. This leads to the modal energy being 
redistributed to greater extents at higher frequencies (Figure~\ref{interact}).
Above the acoustic cut-off frequency, however, the drop in acoustic power makes observing haloes more difficult, requiring the use of other spectral lines or high cadence/resolution observations. 
By choosing Sodium and Potassium lines that form 500 and 250 km above the photosphere, e.g. \citet{moretti} have observed the appearance of haloes at frequencies as high as 7 mHz. 

Hankel analyses of observations \citep[e.g.][]{Braun92,bogdan93,braun95} contain all the features we might expect from this mode mixing theory: (1) the phase shifts increase strongly with frequency, 
(2) the absorption coefficients are seen to initially increase with frequency and saturate at high frequencies for the $p_3$ and higher ridges while for the low mode mass ridges ($p_1, p_2$),
the absorption falls to zero at the high frequencies. Naively, we might conclude that the reason for this drop in the absorption
is an excess of low mode mass waves introduced by scattering from high mode mass ones. However, it turns out that the damping rates of the $p_1,p_2$ modes at high frequency are so 
large that the analyses are dominated by the local excitation and decay of the modes, as opposed arising from correlations across the sunspot \citep[private communication, A. C. Birch 2009; also, Figures 13 and 14 of][]{braun2008}. 
We are therefore unable to conclusively state a case for this theory from just the Hankel observations.

Thus far, we have only discussed the case when the high mode mass waves scatter into the lower ones. A double-edged sword now awaits us: scattering is a two-way street. Equally likely to happen is scattering from low 
mode mass waves into the high, resulting in a surface velocity {\it deficit}. The question then is: does one form of scattering (high to low) preferentially occur over the other? In order to answer this, we
compute the linear interaction of waves with (a) a sunspot sized flux tube, and (b) a bipolar plage-like cluster of moderately sized flux tubes. By ridge filtering the simulated data, we separate the contributions of the 
different $p$ ridges to the halo. We do not present corresponding analyses of the observations because of the low signal-to-noise properties. 

\begin{figure}
\includegraphics[width=\linewidth]{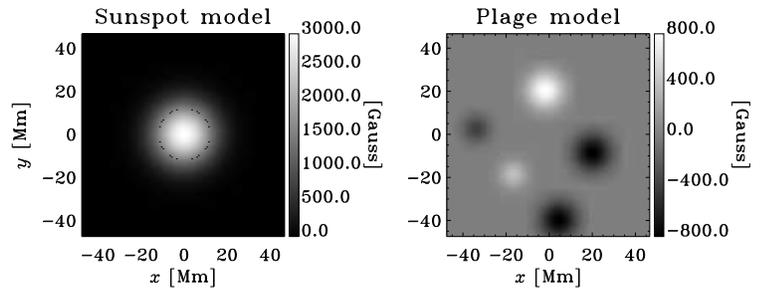}\vspace{0.2cm}
\caption{The magnetic field distributions of the sunspot (left panel) and the bipolar `plage' models (right). The dots mark the location of the half maximum of the field strength in the sunspot model.
\label{magnetogram}}
\end{figure}

The three-dimensional ideal MHD equations are solved via the procedure described in \citet{hanasoge_mag}. For the sunspot calculation, a flux tube with peak photospheric field strength of 3000G
and radius of 10 Mm (full width at half maximum) at the photosphere is placed at the horizontal centre of a box of size $200 \times 200\times 35~{\rm Mm}^3$ (horizontal $\times$ horizontal $\times$ depth)
resolved using $256\times 256\times 300$ grid points. 
The details of the setup are described in \citet{moradi_hanasoge}. The system is deterministically forced with dipolar vertical velocity sources. 
The sources are suppressed in the sunspot region so as to avoid exciting waves in 
 regions of strong magnetic field. In order to determine the extent of the halo effect, we compare the MHD calculation with a quiet simulation, the latter corresponding to a computation of 
 the wave propagation in a domain with sources everywhere and no sunspot. The plage calculation consists of five moderate to weak bipolar flux tubes placed at random locations around the 
 horizontal centre of the simulation box. The plage model used in these calculations fails at mimicking the spaghetti-like complexity of solar plage and we may therefore be significantly 
 underestimating the degree of scatter \citep[e.g.][]{hanasoge_cally}.  Because the plasma-$\beta$ at the source excitation depth is much greater than 1 in the plage case, the sources 
 are chosen to be as in the quiet simulation, i.e., uniform everywhere.  In Figure~\ref{magnetogram}, we display the magnetic field distributions of the sunspot and plage models at a 
 height of 200 km above the photosphere. Vertical velocity components are extracted at an altitude of approximately 200 km above the photosphere (the upper turning point
 of the waves in the quiet simulation is the photosphere). This is done so as to mimic the quiet Sun height of formation of the Nickel line that MDI observes. However, the signals
 are sampled at this geometric height in all regions including in areas of strong magnetic field. We are careful therefore not to interpret the signals in the regions where the field 
 may be significant. It is also important to note that we use a stratification quite different from model S, and consequently the calculations do not contain the same resonant modes as
 the Sun.

Three caveats must be kept in mind when
interpreting the results of these simulations: (1) for the background stratification, we use a combination of sub-surface polytropic and atmospheric isothermal layers, meaning that the resonant frequencies differ from the
solar values, (2) we do not incorporate realistic mode damping, which means the waves are correlated over larger distances, and (3) an Alfv\'{e}n speed limiter is applied in order to prevent the plasma-$\beta$
in the atmosphere from falling too low \citep[e.g.][]{hanasoge_mag} - this avoids negative pressures and lifts severe timestep restrictions. 

We pause the discussion for a moment now to define the terms 
absorption and scattering. The former describes a process by which acoustic waves undergo transformation into slow or Alfv\'{e}n MHD waves \citep[e.g.][]{cally97} and leak out into the atmosphere or disappear
into the interior. They are not a part of the resonant ridges of the observed power spectrum.
The latter denotes mode mixing between the various acoustic modes. In this context, the Alfv\'{e}n speed limiter described above is a cause for some worry since it can alter the balance
between absorption and scattering. Through preliminary work with Hankel transforms (private communication, S. Couvidat, 2008), we have found that the sunspot in our simulations acts more as a
scatterer and less of an absorber. This is to be contrasted with the work of \citet{cameron08}, who find that the sunspots in their simulations absorb very effectively. However the calculations with plage are 
more believable because the plasma-$\beta$ in the tubes become very low only at regions close to the upper boundary. Consequently, the Alfv\'{e}n speed limiter acts only near the boundary and possibly does
not affect the solution within.


\section{Analyses and Conclusions}
\begin{figure}
\includegraphics[width=\linewidth]{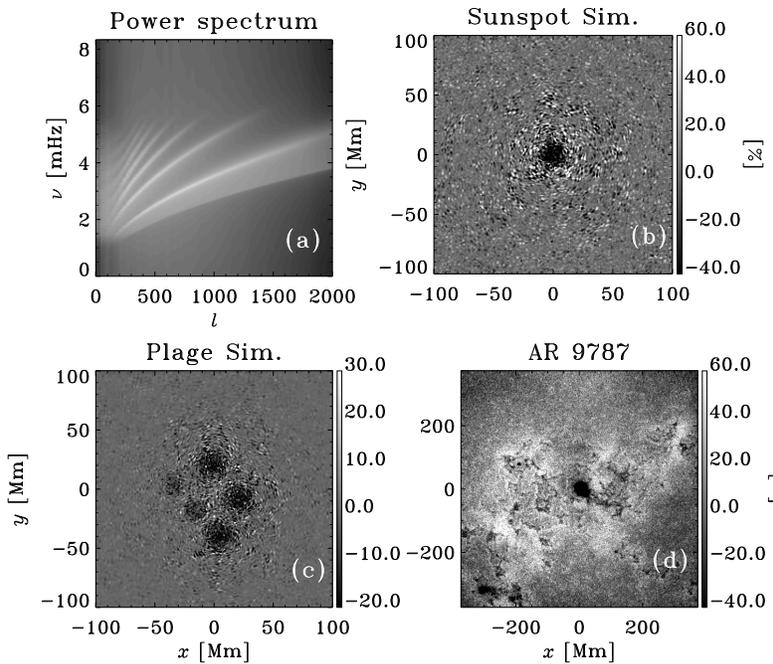}
\caption{Log modal simulation spectrum (panel a) and power maps of the 5-6 mHz frequency-filtered simulated sunspot (panel b), bipolar `plage' (panel c), and full-disk observational data (panel d); the scales are
saturated in all the plots. We do not force the system much below the $f$ ridge, a manifestation of which is the artificial drop in power in panel a. In panels b and c, noise subtraction is applied in order to generate 
a clear view of the acoustic power \citep[e.g.][]{werne04,hanasoge07,hanasoge_mag}. The quiet calculation contains no magnetic field and has a uniform source distribution over the entire horizontal extent of the
domain (i.e., no source suppression in the strong field regions).
\label{powersfreq}}
\end{figure}

The power spectrum of the simulated data is shown in Figure~\ref{powersfreq}a. The high-frequency filtered power maps of the two simulations (panels b,c) and the observations of AR9787 \citep[panel d;][]{scherrer95,gizon2009}
show visible haloes. In order to reduce the noise in the analyses of the simulated data, we also compute a quiet version, i.e. using the same realization of the source function but with no sunspot and sources everywhere. The filtered 
power maps of the quiet cube are then subtracted from the MHD counterparts (see appendix~\ref{filt.app} for a description of the filters). The frequency filtered maps of the simulations in Figure~\ref{powersfreq} (b,c) reproduce 
some of the observational trends - amplitude depressions in the strong field regions and haloes at high frequencies. 

\begin{figure}
\vspace{0.2cm}
\includegraphics[width=\linewidth]{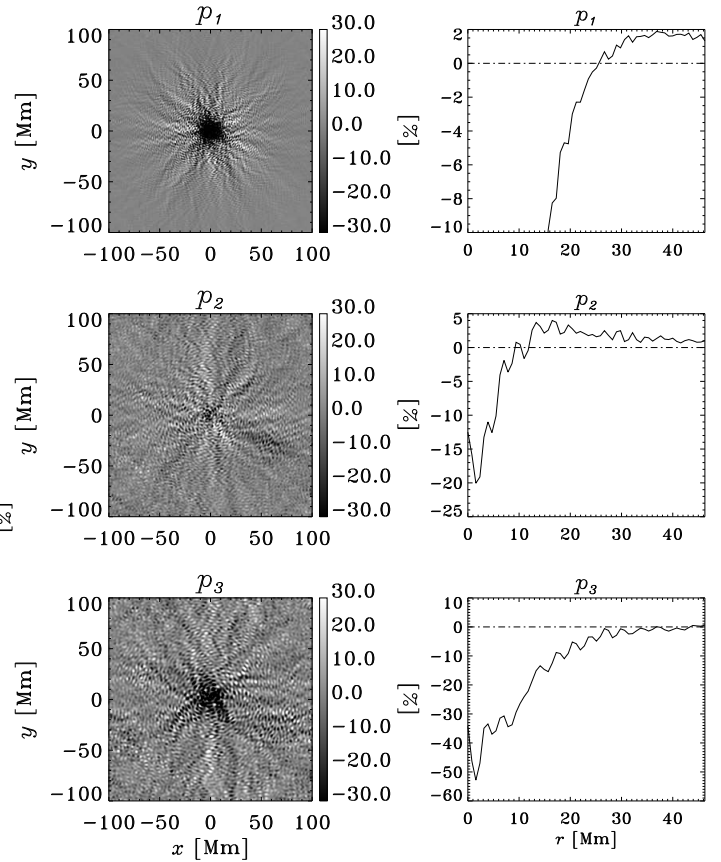}\vspace{0.5cm}
\caption{High-pass (5-6 mHz), $p_1, p_2, p_3$ ridge-filtered power maps of the linear MHD simulations of waves interacting with a sunspot sized flux tube. A halo is seen around the $p_1$ ridge, the $p_2$ map is flooded with 
excess power outside of the strong field regions, and the $p_3$ ridge shows significant power depletion throughout. The panels on the right show the azimuthal averages around the centre of the flux tube. The $y$ range
of the azimuthally averaged $p_1$ power plot has been restricted. Identically filtered power maps of a quiet simulation (i.e. with no sunspot and sources everywhere) have been subtracted
in order to reduce the noise \citep[e.g.][] {hanasoge07,hanasoge_mag}. 
\label{ridgefilters}}
\end{figure}

Shown in Figure~\ref{ridgefilters} are the power maps and flux-tube-centred azimuthal averages of the high-pass- and $p_1, p_2$, and $p_3$ ridge-filtered data. The effect described in the previous section
is seen quite clearly - namely that the $p_1, p_2$ ridge-filtered maps show haloes while a power decrease persists in the $p_3$ map. Strong scattering has caused high mode mass power to be channeled
into the lower mode mass waves.

\begin{figure}
\includegraphics[width=\linewidth]{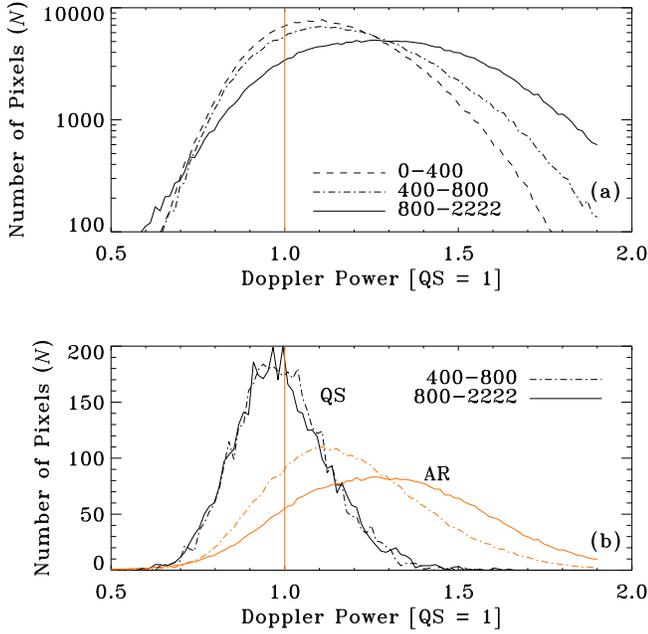}
\caption{High-pass (5-6 mHz) and wavenumber filtered observations of AR 9787: shown in panel a is the number of pixels in the active region power map over which a given relative power level 
persists; in b, we plot the power distributions over the quiet Sun (QS) and active region (AR). The power in each bandpass is normalized with respect to the mean over a correspondingly filtered quiet region; here the quiet Sun 
control is chosen to be a $147\times 61~{\rm Mm}^2$ sized region on the upper left part of Figure~\ref{powersfreq}(d). With increasing wavenumber, larger numbers of pixels display enhancements. The power 
distribution in (b) is also seen to grow progressively asymmetric; the red curves show the active region power distribution in the two bandpasses. 
\label{stats}}
\end{figure}

The most promising demonstration of the validity of this theory is displayed in Figure~\ref{stats}. We take the high-frequency filtered observations of AR9787 and decompose it further into three different spatial bandpasses: $l -
[0,400], [400,800], [800,2222]$, where $l$ is the spherical harmonic degree (see appendix~\ref{filt.app} for a description of the filters). 
Theoretically, we expect to see increasing halo power with spatial bandpass, since at a given frequency, the lowest mode mass waves are at the high wavenumbers. As the quiet Sun control region,
we choose a $\sim 147 \times 61.3 ~{\rm Mm}^2$ sized area, essentially the upper left part of Figure~\ref{powersfreq}(d). The power in each bandpass is normalized by the average power value of the correspondingly filtered quiet 
Sun region. We then compute the areal power distribution: the number of pixels that contain a specific value of relative power are counted and a histogram is plotted. Theoretically, we expect that with growing wavenumber, the 
enhancement area should increase. And indeed, this is exactly what is seen in Figure~\ref{stats}(a)! The number of pixels showing an enhancement greater than 1-$\sigma$, where $\sigma$ is the variance of the power map,
in the three spatial bandpasses is $(0.25, 0.31, 0.38)\times 10^6$ respectively. We also show the quiet Sun and active region power distributions of two bandpasses ([400, 800] and [800,2222]) in panel b of Figure~\ref{stats} in
order to demonstrate that with increasing wavenumber, the distribution shifts in the direction of increasing power.

The theory presented here is capable of explaining the observations of haloes around active regions. The combination of a sunspot and surrounding plage can significantly alter the acoustic
wave power in the vicinity, thereby shifting the local oscillation power spectral density. The magnitudes of the enhancements seen in the simulations are seen to be not dissimilar from the observationally derived values - 
this implies that the mechanism proposed here, with suitable improvements, may be able to reproduce the magnitudes of the observed haloes. 

The analyses of the observations show that the high-wavenumber, high-frequency parts of the spectrum are the most affected, containing the strongest velocity enhancements. The trend that
we sought, and recovered, was that of progressively increasing halo amplitudes with wavenumber. With more sophisticated analyses, we may be able to extract information about mode mixing and perhaps
refine the interpretations of hankel absorption coefficients in light of this.



\begin{acknowledgements}
S.M.H. would like to extend warm thanks to Aaron Birch for the conversations over which this idea appeared and the suggestions that helped improve the paper. 
Thanks also to R. Cameron and L. Gizon for comments. H. Schunker kindly provided the AR9787 observational data. The computing was performed on the Schirra 
supercomputer at NASA Ames. S.M.H. acknowledges support from the German Aerospace Center (DLR) through grant FK Z 50 OL 0801 "German Data Center
for SDO".
\end{acknowledgements}

\appendix
\section{Filters}\label{filt.app}
The frequency filter applied to isolate signals in the 5-6 mHz bandpass is:
\begin{equation}
{\mathcal F}(\nu) = \frac{1}{1 + \exp\left[\frac{0.005-|\nu|}{0.00005}\right]} + \frac{1}{1 + \exp\left[\frac{|\nu| - 0.006}{0.00005}\right]}  -1,\label{freq.filt}
\end{equation}
where $\nu$ is the frequency in Hertz. As for the spatial filtering, we apply, in addition to equation~(\ref{freq.filt}), the following:
\begin{equation}
{\mathcal G}(l) = \frac{1}{1 + \exp\left[l_1 -l\right]} + \frac{1}{1 + \exp\left[l-l_2\right]} -1,\label{space.filter},
\end{equation}
where $l$ is spherical harmonic degree, and $l_1, l_2$ are the lower and upper bounds of the filter, e.g., $(l_1,l_2) = (0,400)$ for the lowest spatial bandpass of Figure~\ref{stats}(a).


\bibliographystyle{aa}
\bibliography{Halos}
\end{document}